\documentclass[aps,prb,twocolumn,showpacs,superscriptaddress,floatfix]{revtex4-1}
\usepackage{graphics}
\usepackage{graphicx}
\usepackage{dcolumn}
\usepackage{bm}
\usepackage{epsfig}
\usepackage{float}
\newcommand{\degree}{\ensuremath{^\circ}}

\begin{document}

\title{Electronic structure of KTi(SO$_{4}$)$_{2}\cdot$H$_{2}$O - a S=1/2 frustrated chain antiferromagnet }

\author{Deepa Kasinathan}
\email{Deepa.Kasinathan@cpfs.mpg.de}
\affiliation{Max Planck Institut f\"{u}r Chemische Physik fester Stoffe,
N\"{o}thnitzer Str. 40, 01187 Dresden, Germany}

\author{K. Koepernik}
\affiliation{IFW Dresden, P.O. Box 270116, D-01171 Dresden, Germany}
\author{O. Janson}
\affiliation{Max Planck Institut f\"{u}r Chemische Physik fester Stoffe,
N\"{o}thnitzer Str. 40, 01187 Dresden, Germany}

\author{G. J. Nilsen}
\affiliation{Laboratory for Quantum Magnetism, ICMP, EPFL, CH-1015, Lausanne, Switzerland}

\author{J. O. Piatek}
\affiliation{Laboratory for Quantum Magnetism, ICMP, EPFL, CH-1015, Lausanne, Switzerland}

\author{H. M. R$\o{}$nnow}
\affiliation{Laboratory for Quantum Magnetism, ICMP, EPFL, CH-1015, Lausanne, Switzerland}

\author{H. Rosner}
\email{rosner@cpfs.mpg.de}
\affiliation{Max Planck Institut f\"{u}r Chemische Physik fester Stoffe,
N\"{o}thnitzer Str. 40, 01187 Dresden, Germany}

\date{\today }

\begin{abstract}

The compound KTi(SO$_{4}$)$_{2}\cdot$H$_{2}$O was recently reported as
a quasi one-dimensional spin 1/2 compound with competing
antiferromagnetic nearest neighbor exchange $J_1$ and next-nearest
neighbor exchange $J_2$ along the chain with a frustration ratio
$\alpha$ = $J_{2}/J_{1}\approx$ 0.29 [Chem. Mater. {\bf 20}, 8 (2008)].  Here, we report a
microscopically based magnetic model for this compound derived from
density functional electronic structure calculations along with
respective tight-binding models. Our calculations confirm the quasi
one-dimensional nature of the system with antiferromagnetic $J_{1}$
and $J_{2}$, but suggest a significantly larger frustration ratio
$\alpha \approx 1.1 \pm 0.2$.
Based on transfer matrix renormalization group calculations we found that, due to an intrinsic symmetry
of the $J_1$-$J_2$ model, our larger frustration ratio $\alpha$ is
also consistent with the previous thermodynamic data. 
To resolve this issue, we propose performing high-field magnetization measurements 
and low temperature susceptibility measurements which should allow 
to precisely identify the frustration ratio $\alpha$.

\end{abstract}

\pacs{look it up}

\maketitle
\section{Introduction}

For several decades, low dimensional magnetism has attracted great interest in
solid state physics and chemistry. Since the influence of quantum
fluctuations becomes especially pronounced for low dimensional
spin-1/2 systems, these systems have been investigated extensively,
both theoretically and experimentally.  Quantum fluctuations become
even more important in determining the ground state and the nature of the low lying excitations
if the system under consideration exhibits strongly frustrated
interactions.

Whereas pure geometrical frustration, i.e triangular, Kagom\'e or
pyrochlore lattices, can be realized by special symmetries in two or
more dimensions, frustration in one-dimensional systems (linear chains) is 
generally realized by competing interactions. The simplest frustrated one
dimensional model is the $J_1$-$J_2$ model with nearest neighbor (NN)
exchange $J_1$ and next nearest neighbor (NNN) exchange $J_2$, where 
$J_2$ is antiferromagnetic. The
phase diagram of this seemingly simple model is very rich. Depending
on the frustration ratio $\alpha = J_2/J_1$, a variety of ground
states are observed in corresponding quasi-1D systems: (i)
ferromagnetically ordered chains in Li$_2$CuO$_2$ 
\cite{DrechslerEPL,Nitzsche} ($0 \leq |\alpha| \leq 0.25$); 
(ii) helical order with different
pitch angles in LiVCuO$_4$, LiCu$_2$O$_2$, and NaCu$_2$O$_2$
\cite{Enderle,Masuda,Gippi04,Drechs05,Capogna,Drechs06,LiCu2O2} 
($\alpha < -0.25$); (iii) spin-Peierls transition in
CuGeO$_3$ \cite{CuGeO3,hase} ($\alpha \gtrsim 0.2411$), i.e.~both exchange
couplings are antiferromagnetic.

The possibility of subtle interplay between spin, orbital, charge and lattice 
degrees of freedom due to the threefold orbital degeneracy of Ti$^{3+}$
in octahedral environments, makes titanium 3+ based 
oxides an interesting class of materials to study. Exotic features like 
orbital-liquid state in LaTiO$_{3}$ and presence of strong orbital 
fluctuations in YTiO$_{3}$ have been reported.\cite{LaTiO3,YTiO3} 
While an abundance of experimental work exists on low dimensional spin-1/2
cuprates (with Cu$^{2+}$), materials based on spin-1/2
titanates (with Ti$^{3+}$) are rather sparse. 
A well known example of low-dimensional titanates is the new class
of inorganic spin-Peierls materials TiO$X$ ($X$ = Cl, Br).\cite{seidel}
Quasi one-dimensional magnetism was observed in TiOCl and TiOBr
 along with a first-order transition to a dimerized non-magnetic ground
 state (spin-Peierls like) below 67 K and 27 K, respectively. 
The main difference between S=1/2 titanates and vanadates
as compared to the cuprates (3$d^1$ vs. 3$d^9$) is that the unpaired
electron resides in the $t_{2g}$ complex for the former, while
occupying the $e_{g}$ complex for the latter. This usually results in
narrow bands at the Fermi level (E$_{F}$) for the titanates and
vanadates as compared to the cuprates, which in turn leads to
small values for the exchange couplings and brings
several experimental conveniences. Since the temperature scale for
the magnetic contribution to the specific heat is of the order of $J$,
separating magnetic and phonon contribution in the specific heat
($C_p$) measurements is thus relatively easy.  For the same reason,
magnetization measurements can reach the saturation moment in
experimentally attainable fields, thereby providing more information
about the exchange parameters and the frustration regime.  Thus, in
many respects, spin 1/2 compounds with 
a singly occupied $t_{2g}$ orbital
are an ideal object to study the 
physics of quasi one-dimensional frustrated chains.

\begin{figure}[t]
\begin{center}
\includegraphics[%
  clip,
  width=8cm,
  angle=-0]{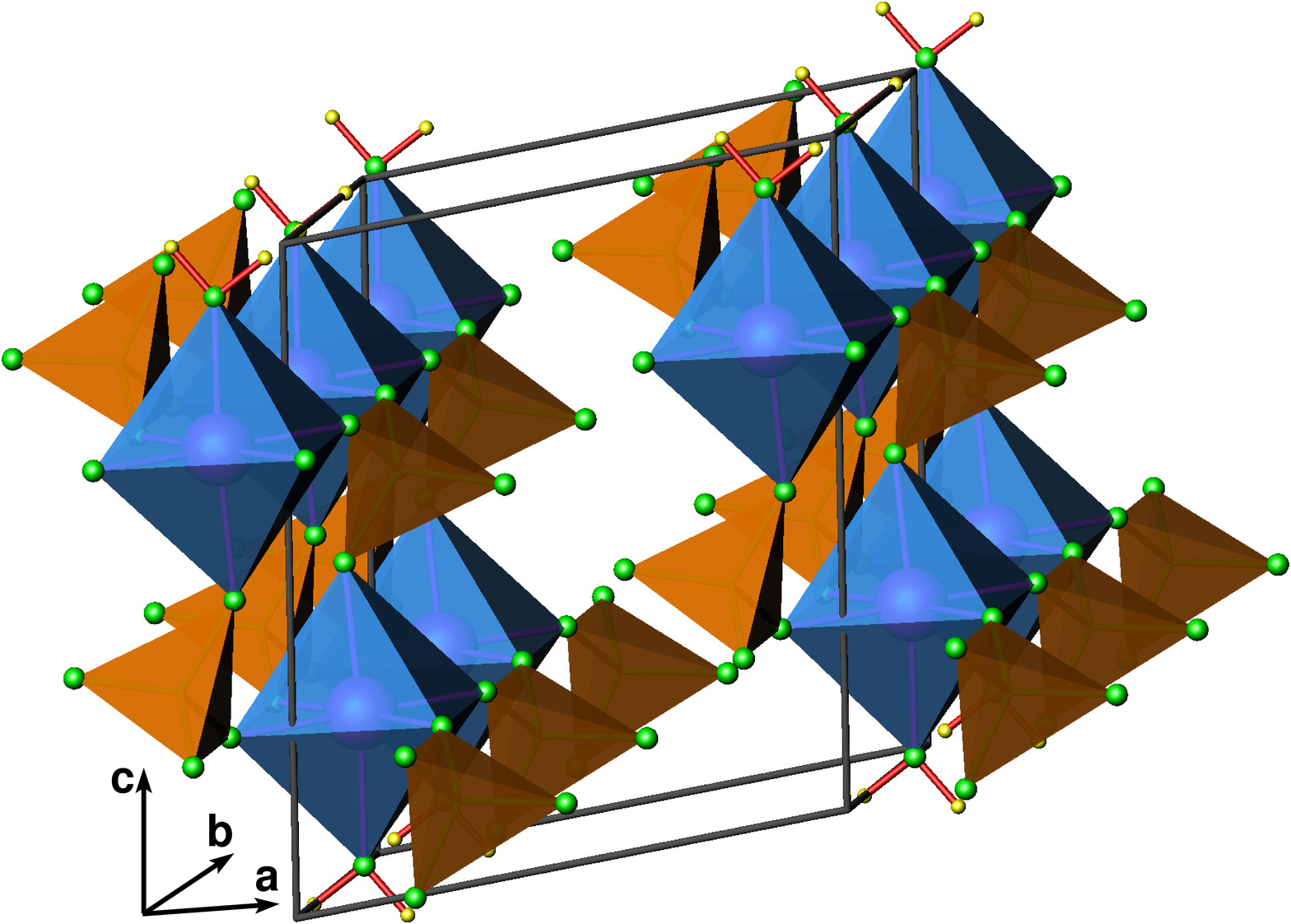}
\includegraphics[%
  clip,
  width=5cm,
  angle=-0]{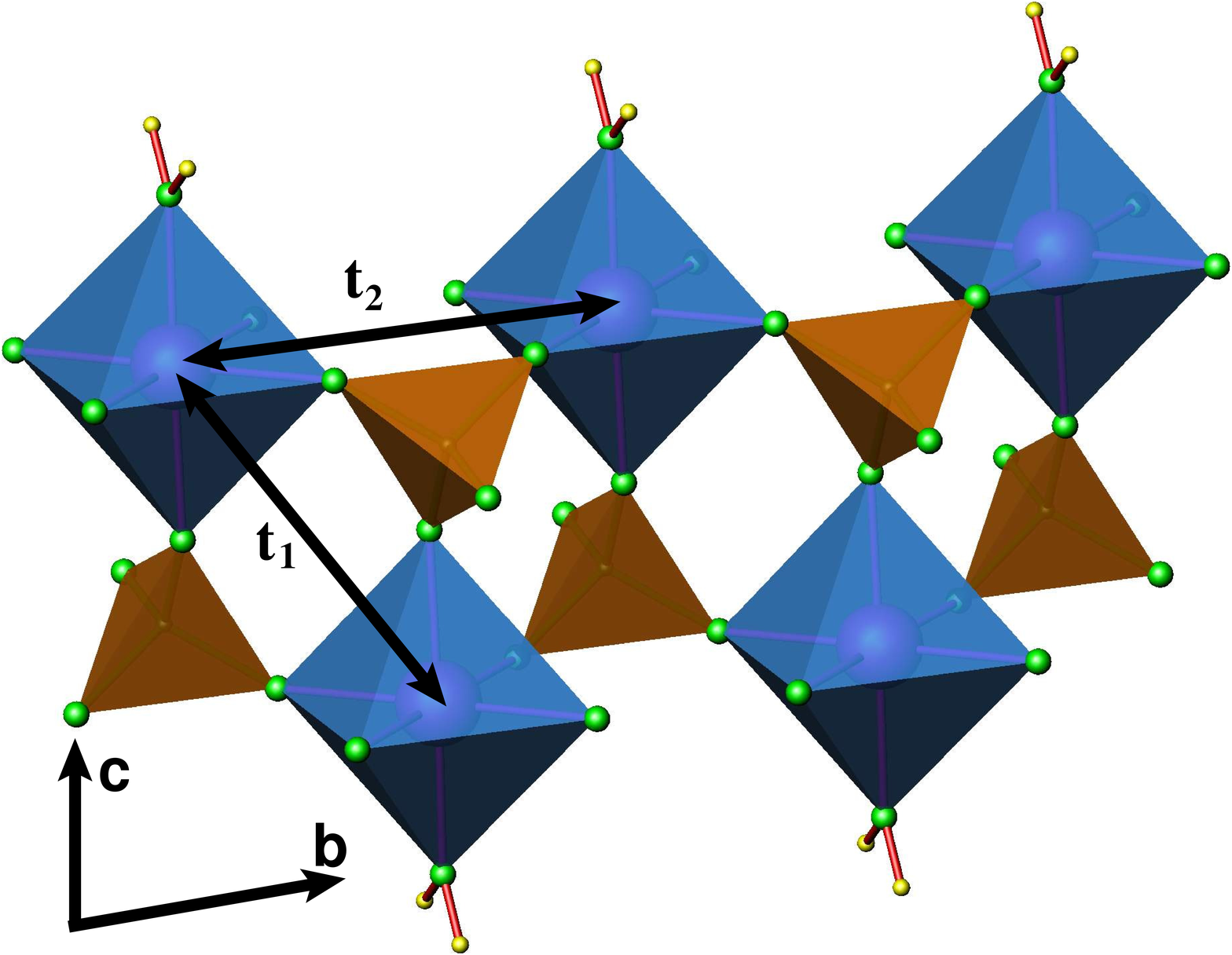}
\end{center}
\caption{\label{str}(Color Online) {\bf Top:} Crystal structure of 
KTi(SO$_{4}$)$_{2}\cdot$H$_{2}$O. Double chains of TiO$_{6}$ octahedra
run along the $b$-axis. The octahedra are connected on either sides by
SO$_{4}$ tetrahedra. Water molecules are bound to the octahedra and
separating the double chains along the $c$-axis. The potassium atoms
(not shown here) separate the double chains along the
$a$-axis. {\bf Bottom:} An isolated segment of the double chain. The
nearest neighbor (NN) and next nearest neighbor (NNN) hopping paths
are represented as t$_{1}$ and t$_{2}$ respectively. }
\end{figure}

KTi(SO$_{4}$)$_{4}\cdot$H$_{2}$O, a new member in the family of titanium
alums has recently been
synthesized\cite{nilsen08}. The titanium ions are in the 3+ (d$^{1}$)
oxidation state in this material. Specific heat and susceptibility measurements
indicate that KTi(SO$_{4}$)$_{4}\cdot$H$_{2}$O is a S=1/2 frustrated chain system. 
Fits to the susceptibility data using exact diagonalization on up to 
18 spins resulted in estimates for 
leading exchanges $J_{1}$ = 9.46~K and $J_{2}$ = 2.8~K, both AFM
with $\alpha$ = 0.29.
In the well studied $J_{1}$-$J_{2}$ frustrated chain model with both interactions 
being AFM, the model undergoes a quantum phase transition at
$\alpha$ = 0.2411 to a two-fold degenerate gapped phase, and for $\alpha >$ 0.2411 the system
will exhibit spontaneous dimerization.\cite{alpha1,alpha2,alpha3,alpha4,alpha5,
alpha6,alpha7} 
In light of this result, KTi(SO$_{4}$)$_{4}\cdot$H$_{2}$O would be expected to occupy
the highly interesting region of the phase diagram,
with a small gap $\Delta < J_{1}$/20.\cite{alpha7,nilsen08}

Here, we report the results of an
electronic structure analysis from first principles, carried out to obtain a
microscopic picture of the origin of the low dimensional magnetism in
KTi(SO$_{4}$)$_{4}\cdot$H$_{2}$O. 
The magnetically active orbital is identified using band structure 
calculations followed by subsequent analysis of the exchange couplings. 
The calculated $J$'s have the same
sign as the experiments ($i.e.$ AFM), but the estimated frustration ratio $\alpha$ is 
considerably larger compared to the experimental findings. We explore in 
detail the reasons
for this  discrepancy and propose an alternative solution that fits
the experimental data, as well as being consistent with our band structure
calculations. To this end, we have 
simulated the temperature dependence of
magnetic susceptibility
using the transfer-matrix renormalization group (TMRG) method to 
unambiguously identify the exchange couplings that describe the microscopic
magnetic model for KTi(SO$_{4}$)$_{2}\cdot$H$_{2}$O. 
On the basis of these results, we suggest performing high-field magnetization measurements,
which will be a decisive experiment to identify the precise frustration ratio $\alpha$.
The influence of crystal water on the observed ground
state is also discussed in detail.

The remainder of the manuscript is organized as follows: The crystal structure 
of KTi(SO$_{4}$)$_{2}\cdot$H$_{2}$O is described in section II. The details 
of the various calculational methods are collected in section III. 
The results of the density functional theory based calculations, including
the band structure and the accordingly derived microscopic model is 
described in section IV. In section V we demonstrate the effects of the
crystal water in this compound by calculating the Wannier functions. 
The outcome of the TMRG simulations is compared to the previous
experiment in section VI, which is followed by a discussion and summary in 
section VII. 

\section{Crystal Structure}

Throughout our calculations, we have used the recently determined
experimental\cite{nilsen08} lattice parameters in the monoclinic
space group ($P2_{1}/m$) of KTi(SO$_{4}$)$_{2}\cdot$H$_{2}$O:
$a$~=~7.649~\AA, $b$~=~5.258~\AA, $c$~=~9.049~\AA~and
$\beta$~=~101.742$^{\degree}$. The crystal structure which is
isomorphous to that of the mineral Krausite\cite{krausite},
KFe(SO$_{4}$)$_{2}\cdot$H$_{2}$O is displayed in
Fig.~\ref{str}. Isolated pairs of chains (``double chains") of
TiO$_{6}$ octahedra run along the crystallographic $b$-axis. The
octahedra are distorted and have no edge sharing or corner sharing
oxygen atoms. The SO$_{4}$ tetrahedra 
corner share with three adjacent TiO$_{6}$ octahedra, forming an isosceles
triangle. These trangles edge-share to make up the double chains.
The single chains are displaced with respect to
each other both laterally and vertically. Large K$^{+}$ ions isolate
the double chains along $a$, while along $c$ the chains are separated by
water molecules that share the oxygen atom with the TiO$_{6}$
octahedra. The water molecules are oriented in the $ac$ plane. 
All of the octahedral O-Ti-O bond angles deviate slightly
away from 90$^{\degree}$ and there are three pairs of Ti-O bond lengths of
2.001~\AA, 2.056~\AA~ (in the $ab$ plane) and 2.043~\AA~ (along $c$).
The shortest Ti-Ti distance is 4.93~\AA, and is between nearest neighbors (NN) on
the adjacent chains ($t_{\mathrm{1}}$ in Fig.~\ref{str}). 
Within the chains, neighboring Ti are 5.23~\AA~apart($t_{\mathrm{2}}$ in Fig.~\ref{str}). 
By analogy 
with the $J_{\mathrm{1}}-J_{\mathrm{2}}$ Heisenberg model, we will call the  
magnetic interactions corresponding to the shorter distance (NN)
$J_{\mathrm{1}}$ and the longer distance (next nearest neighbors, NNN) $J_{\mathrm{2}}$.
In case of a perfect
octahedral environment, the three $t_{2g}$ states would be degenerate
and thus warrant additional effects ($i.e.$ lattice distortion,
spin/charge/orbital ordering) to lift the degeneracy and allow for
an S=1/2 singlet ground state for a Ti$^{3+}$ ion.\cite{fn}  In
KTi(SO$_{4}$)$_{2}\cdot$H$_{2}$O, there is a small distortion of the
octahedra and consequently splitting of the $t_{2g}$ levels
can be expected.  The related case of TiOCl, another system containing
Ti$^{3+}$ ion, also possesses a distorted arrangement of TiCl$_{2}$O$_{4}$
octahedra, though the distortions are much larger with 
equatorial Ti-O and Ti-Cl bond lengths of 2.25~\AA~ and 2.32~\AA, respectively,
and an  apical Ti-O bond length of 1.95~\AA. 
Consequently, the $t_{2g}$ orbitals in TiOCl were thought to split into a lower energy
$d_{xy}$ and higher energy $d_{xz,yz}$ orbitals. Electronic structure
calculations confirmed this interpretation and revealed the magnetically active orbital for the S=1/2
chains in TiOCl was indeed the lower energy $d_{xy}$
orbital,\cite{seidel,saha} though a prolonged discussion of possible orbital 
fluctuations ensued afterwards. Therefore, an analysis of KTi(SO$_{4}$)$_{2}\cdot$H$_{2}$O 
from the structural point
of view alone is not sufficient to determine the ground state of the
system. Detailed calculations are necessary to understand the correct
orbital and magnetic ground state of the system.  
 
\section{Calculational Details}

The DFT calculations were performed using a full potential 
nonorthogonal local orbital code (\texttt{FPLO}) within the local (spin)
density
approximation (L(S)DA) \cite{fplo1,fplo2}. 
The energies were converged on a dense $k$ mesh with 
300 points for the conventional cell in the irreducible wedge of the Brillouin
zone. The Perdew and Wang flavor\cite{PW92}of the
exchange
correlation potential was chosen for the scalar relativistic calculations.
The strong on-site Coulomb repulsion of the Ti 3$d$ orbital was taken into 
account using the L(S)DA+$U$ method, applying the ``atomic limit'' double
counting term. The projector on the
correlated orbitals was defined such that the trace of the occupation number 
matrices represent the 3$d$ gross occupation.
 The AFM parts of the exchange couplings are computed as 
 $J_{i}^{\mathrm{AFM}} = 4t_{i}^{2}/U_{\mathrm{eff}}$,  by mapping the
 results of the LDA calculations on to a tight binding model (TBM) which 
 is then mapped on to a Hubbard model, and subsequently to a Heisenberg 
 model because the system belongs to
the strong correlation limit $U_{\mathrm{eff}}$ $\gg$ $t$ ($t$ is the leading transfer 
integral at half filling). The full exchange couplings are obtained by mapping the LSDA+$U_{d}$ total energies 
 of various supercells with collinear spin configurations to a classical Heisenberg 
 model. The supercells used to calculate $J_{\mathrm{1}}$ and $J_{\mathrm{2}}$ 
 were sampled using 300 and 100 $k$ points, respectively. 
 Maximally localized Wannier functions (WF) were calculated for the 
Ti $t_{2g}$ orbitals, also using \texttt{FPLO}~\cite{wanfun}, to obtain a 
visual insight of the relevant orbitals and superexchange paths. 

The magnetic excitation spectrum of frustrated spin chains was simulated
using exact diagonalization code from the ALPS package.\cite{ALPS} We
used periodic boundary conditions and considered finite lattices 
comprising up to $N\,=\,32$ spins.

The magnetic susceptibility of infinite frustrated spin chains was simulated
using the transfer-matrix renormalization group (TMRG)
technique.\cite{tmrg} For each simulation, we kept 120--160 states, the
starting inverse temperature was set to 0.05$J_1$, and the Trotter
number was varied between 4$\cdot$10$^3$ and 16$\cdot$10$^3$. The
results were well converged for the whole temperature range of the
experimental curve from Ref.~\onlinecite{nilsen08}.

\begin{figure}[t]
\begin{center}
\includegraphics[%
  clip,
  width=8cm,
  angle=-0]{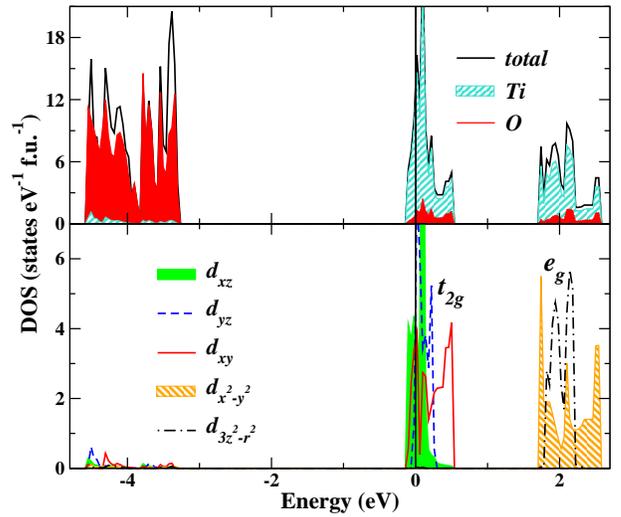}
\end{center}
\caption{\label{dos}(Color Online) Top: Total and partial DOS obtained within
LDA for KTi(SO$_{4})_{2}\cdot$H$_{2}$O. The valence panel is predominantly
comprised of
Ti-3$d$ and O-2$p$ states. The sulphur states (not shown here) lie mainly below -8 eV.
The contribution from K, S and H sites are negligible
in the displayed energy range. Bottom: Ti 3$d$-orbital resolved DOS. The
$t_{2g}$ and $e_{g}$ complexes are split by a ligand field splitting of 
about 2~eV. 
The $d_{xz}$ orbital and the $d_{yz}$ orbital are very close and split from
the broad $d_{xy}$ (larger bandwidth) orbital. } 
\end{figure}

\section{Electronic Structure Calculations}

\subsection{Local Density Approximation}
Since there exists no previous report on the electronic structure of
KTi(SO$_{4}$)$_{2}\cdot$H$_{2}$O, we begin by analyzing the results
from a non-magentic LDA calculation. 
In a simplified, fully ionic model each Ti$^{3+}$ ion is surrounded by 
a slightly distorted octahedron of O$^{2-}$ ions.
The 
pseudo octahedral coordination dictates a set of local axes for the
conventional $e_{g}$ and $t_{2g}$ orbitals. 
The local coordinate system is chosen as $\hat{z}~||~ c$, and 
$\hat{x}$ and $\hat{y}$ axes are rotated by 45$^{\degree}$ around $c$ with respect
to the original $a$ and $b$ axes. 
The non-magnetic total and orbital resolved density
of states (DOS) are collected in Fig.~\ref{dos}. 
The presented part of the valence band is predominantly
comprised of Ti 3$d$ and O 2$p$ states belonging to TiO$_{6}$ octahedra.
The states belonging to sulfur (not shown) lie below 
-8 eV and are therefore well separated from the TiO$_{6}$ states. 
The weight close to the Fermi level
(E$_{F}$) is mainly from the Ti $t_{2g}$ states which contain 
two electrons (one for each Ti in the unit cell) and are
separated by a ligand field splitting of about 2~eV from the higher lying 
(empty) 
$e_{g}$ states. For an octahedral arrangement of oxygen 
anions around a 3$d$ transition metal cation, a 2~eV ligand field
split is rather typical. 
For KTi(SO$_{4}$)$_{2}\cdot$H$_{2}$O, 
the band width of the $t_{2g}$ band complex is only 0.65~eV, 
about one third of the value for TiOCl ($\sim$2~eV)\cite{saha}. 
This difference arises from the fact that in TiOCl, the basic octahedral
structural units TiCl$_{2}$O$_{4}$ are arranged such that
they are corner-sharing in the $a$-direction and edge-sharing along
the $b$-direction, leading to a larger interaction between the octahedral
units and hence a larger $t_{2g}$ band width. In contrast, the TiO$_{6}$ octahedral 
units in KTi(SO$_{4}$)$_{2}\cdot$H$_{2}$O are neither corner- nor edge-sharing
and hence the smaller $t_{2g}$ band width.  

\begin{figure}[t]
\begin{center}
\includegraphics[%
  clip,
  width=8cm,
  angle=-0]{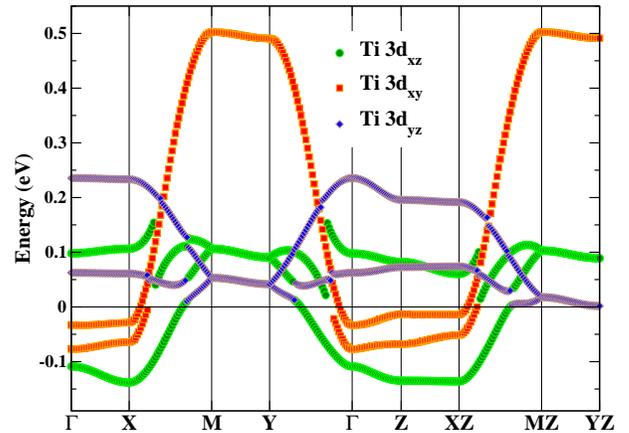}
\end{center}
\caption{\label{bands}(Color Online) Band structures with band character 
close to the Fermi level obtained within LDA. The bands are dispersive 
along $\Gamma$-Y and X-M direction - the crystallographic $y$ direction
along which the ``double chains'' propagate. There are two Ti atoms in the
unit cell and therefore 6 $t_{2g}$ bands crossing the Fermi energy. 
} 
\end{figure}

The degeneracy between the $t_{2g}$ orbitals is lifted due to the monoclinic
symmetry of the crystal structure as seen in the non-magnetic band structure
(Fig.~\ref{bands}). There are 2 Ti atoms per formula unit and therefore 
6 $t_{2g}$ bands close to E$_{F}$. 
The bands are highly dispersive along $\Gamma$-Y, X-M and XZ-MZ directions, 
which are along the crystallographic $y$-axis and remain rather flat along 
the other high symmetry directions. This implies that the main interaction
between the Ti$^{3+}$ ions is along the ``double chain'', while sizably smaller 
interactions are expected between the adjacent double chains. 
The band belonging to the $d_{xz}$ orbital is lower in energy as compared 
to the $d_{yz}$ 
and $d_{xz}$ (nearly empty and larger band width) orbitals (also see Fig.~\ref{dos}). 
The mixing between the Ti 3$d$ and the O 2$p$ states 
close to E$_{F}$ is less than 10\% and similar to other systems
where Ti occurs in the $d^{1}$ configuration. 
This scenario is different from the
cuprates (d$^{9}$) where 30\% of the contribution to the states at 
E$_{F}$ stems from O 2$p$. This fundamental difference in the strength of
the hybridization between the transition metal ions and the oxygen ligands
comes from the relative energies ($\Delta$) of the oxygen $p$ and the 
transition metal $d$ bands. In cuprates, the highest (half-filled) 
$d_{x^{2}-y^{2}}$ orbital and the uppermost (filled) oxygen $p$-orbitals 
are rather close in energy ($\Delta \sim$ 2~eV) resulting in a strong $pd$
hybridization. In titanates, on the contrary, the $t_{2g}$ orbitals lie
much higher in energy than the uppermost (filled)
oxygen $p$-orbitals ($\Delta \geq$ 3~eV) and therefore exhibit
significantly less $pd$ 
hybridization. Upon hole doping, the holes would formally appear in the oxygen
$p$-orbitals for cuprates and in one of the $t_{2g}$ orbitals for titanates,
characterizing them as charge-transfer and Mott-Hubbard systems, respectively. 

Experimentally, KTi(SO$_{4})_{2}\cdot$H$_{2}$O
is an insulator, but a metallic solution is obtained within LDA.
Such metallic results from LDA are well known and understood to arise
from the inadequate treatment of the strong Coulomb correlation of the 
3$d$ orbitals. Therefore, the orbital dependence of the Coulomb and exchange 
interactions are taken into account in a mean field like approximation 
using the LSDA+$U$ approach (Section IV-C).
As mentioned previously in Sec. II, the distorted octahedra in TiOCl split the 
$t_{2g}$ states into a lower lying singlet ($d_{xy}$) and a higher energy
doublet ($d_{xz,yz}$). Presuming no further symmetry breaking, adding 
correlations, the choice 
of the orbital for occupying the one 
unpaired electron of the Ti$^{3+}$ ion in TiOCl is rather straightforward:
the singlet $d_{xy}$. On the other hand, for KTi(SO$_{4})_{2}\cdot$H$_{2}$O
the pseudo octahedral ligand field splits the $t_{2g}$ states and removes the
three fold degeneracy. Nevertheless, due to the relatively small energy differences, 
there is no clear candidate for the 
choice of the half-filled orbital. 
The band center of the $d_{xz}$ orbital is around E$_{F}$ and is the
lowest lying band of the $t_{2g}$ complex. The center of gravity of 
$d_{yz}$ and the comparably 
broad $d_{xy}$ band are only
slightly higher in energy than the $d_{xz}$ at 0.15 eV and 0.25 eV, respectively. 
Since the three 
$t_{2g}$ orbitals are quite close in energy, a subtle balance between the 
orbitals is expected and therefore one should carefully consider the possibility
of an orbitally ordered solution. At this juncture, it is rather unclear
which correlated orbital will be half filled and thereby determine the 
magnetic model. Even the undoped, low dimensional S=1/2 cuprates that possess an extensive literature,
where the magnetic model is generally understood to be governed by the 
half-filled $d_{x^{2}-y^{2}}$
orbital, can sometimes show surprises. For example, the CuO$_{6}$
octahedral environment in the insulating S=1/2 quasi 1D system
CuSb$_{2}$O$_{6}$ is less distorted than usual, so that the cubic degeneracy
for the $e_{g}$ 
ligand-field states are only slightly lifted. The
energy difference between the $d_{x^{2}-y^{2}}$ and $d_{3z^{2}-r^{2}}$ related
band centers is about 0.3 eV only, compared with
about 2 eV for standard cuprates.
Inclusion of correlations, changes the order and 
results in the 
unpaired electron occupying the $d_{3z^{2}-r^{2}}$ orbital instead of the 
standard $d_{x^{2}-y^{2}}$ in CuSb$_{2}$O$_{6}$.\cite{CuSbO}   
 
We take into account the strong
electronic correlations of the Ti 3$d$ states via two possible ways:
(a) mapping the results from LDA first to a tight-binding model (TBM),
which in turn is mapped onto a Hubbard
model, and subsequently to a Heisenberg model.
At half filling, when $U$ is much larger than the bandwidth $w$,
the spin degrees of freedom are well described by an S=1/2 AFM Heisenberg 
Hamiltonian with an exchange interaction $J_{i}^{\text{AFM}} \simeq 4t_{i}^{2}/U_{\text{eff}}$.
(b) performing LSDA+$U_{d}$ total energy calculations for various 
collinear spin configurations and mapping the energy differences on to 
a classical Heisenberg model to obtain the total exchange $J$. 
At this juncture, the difference between the two parameters used to incorporate the
effects of strong correlations, $U_{\mathrm{eff}}$ and 
$U_{d}$ must be clarified. The former is applied to LDA bands which include the 
effects of hybridization between the metal atoms and ligands, while the latter is 
applied to purely atomic $d$-orbitals and therefore necessitates using different values 
for these two parameters.  

\begin{center}
\begin{table}[t]
\begin{tabular*}{0.47\textwidth}%
     {@{\extracolsep{\fill}}|c|c|c|c|c|c|}
\hline
 & & & & & \tabularnewline
 & t$_{1}$ (eV) & t$_{2}$ (eV) & $J_{1}^{\mathrm{AFM}}$ (K) & $J_{2}^{\mathrm{AFM}}$ (K) &
$\alpha$ \tabularnewline
 & & & & & \tabularnewline
\hline
$d_{xz}$ & 0.047 & 0.029 & 31 & 11.8  & 0.38 \tabularnewline
$d_{yz}$ & 0.037 & 0.028 & 19 & 11 & 0.57 \tabularnewline
$d_{xy}$ & 0.081 & 0.135 & 92 & 256 & 2.78 \tabularnewline
\hline
\end{tabular*}
\caption{\label{tbm} Hopping parameters (in meV) and the corresponding 
exchange constants (in K) from an effective one-band TBM. The hopping paths
are indicated in the lower panel of Fig.~\ref{str}. 
The transfer integral $t_{1}$ corresponds to hopping between the single
chanis, while $t_{2}$ corresponds to hopping within a single chain. 
A $U_{\mathrm{eff}}$ of 3.3 eV has been used to calculate the $J$'s.
The frustration 
ratio $\alpha$ = $J_{2}$/$J_{1}$ is collected in the last column.  }
\end{table}
\end{center}

\subsection{Tight Binding Model}
Though LDA fails in reproducing the insulating ground state of KTi(SO$_{4}$)$_{2}\cdot$H$_{2}$O
system, it still provides valuable information about the orbitals involved
in the low energy physics, as well as their 
corresponding interactions strengths. 
As mentioned previously, the strong Coulomb correlations favor full 
polarization and a detailed analysis is necessary to identify the ground
state
for the fully polarized $d$-orbital.  
Therefore, to obtain a microscopic picture of the magnetic interactions,
we have first derived an effective one-band TBM
for each set of the $t_{2g}$ bands. 
For each set of the different ``active'' $t_{2g}$ orbitals, we fit the 
corresponding LDA bands to a separate TBM. 
In second quantization formalism, the tight-binding Hamiltonian can be expressed as,
\begin{equation}
 \hat{H}_{\text{TB}} =  \sum_{\langle i,j \rangle,\sigma} t_{ij} (\hat{c}^{\dagger}_{i\sigma}\hat{c}_{j\sigma} + \hat{c}^{\dagger}_{j\sigma}\hat{c}_{i\sigma} ) + \sum_{i,\sigma} \epsilon_{i} \hat{c}^{\dagger}_{i\sigma}\hat{c}_{i\sigma} 
\end{equation}
where $c^{\dagger}_{i\sigma}, c_{j\sigma}$ are the usual creation and annihilation 
operators; $\sigma$ denotes the spin polarization; $\epsilon_{i}$ are the constant 
energy shifts with respect to the Fermi level; $t_{ij}$ are the transfer integrals and
$\langle i,j \rangle$ are the site indices.
The magnitudes of the leading hopping integrals (the paths are indicated in
Fig.~\ref{str}) are collected in Table~\ref{tbm}. 
The NN (between single chains) and NNN (along the single chains) hopping 
integrals $t_{1}$ and $t_{2}$ are of the same order of magnitude
for the $d_{xz}$ and $d_{yz}$ orbitals. On the contrary, the TBM fit to the broad 
$d_{xy}$ orbital leads to a much larger $t_2$. 
All other $t$'s beyond NNN are less than 0.1 meV  for all the 
three orbitals and can be neglected for the chain physics. The two main hopping terms are 
thus confined to interactions between the Ti sites within each S = 1/2
``double chain'',
consistent with the experimental observations\cite{nilsen08} of 
displaying low dimensional magnetic properties.
The individual exchange
constants are then calculated using 
$J_{i}^{\mathrm{AFM}}$ = 4t$_{i}^{2}$/$U_{\mathrm{eff}}$. 
For TiOCl, a $U_{\mathrm{eff}} \sim$ 3.3 eV was shown to provide good
agreement between calculated exchange constants and susceptibility 
measurements\cite{saha}. The same value of 
$U_{\mathrm{eff}}$ = 3.3 eV has been used here for
KTi(SO$_{4}$)$_{2}\cdot$H$_{2}$O, the $J$'s obtained
are collected in Table.~\ref{tbm}. 
The $J$'s for the lower energy $d_{xz}$ band and as well as the $d_{yz}$ band 
are of similar magnitude as compared
to the experimental report. To the contrary, both the calculated NN and NNN
$J$'s for the $d_{xy}$ band are much larger in energy compared to the experimental report.    
This large difference in the energy scale of the magnetic exchange for 
$d_{xy}$ band implies that this orbital might be a rather
unlikely choice for full polarization (a more clear picture emerges in the
following section when performing LSDA+$U$ calculations).

\begin{figure}[t]
\begin{center}
\includegraphics[%
  clip,
  width=8cm,
  angle=-0]{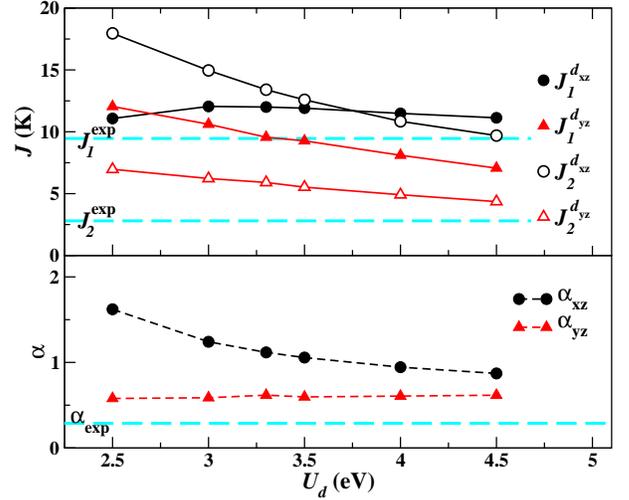}
\end{center}
\caption{\label{alpha}(Color Online) The total exchange constants $J$'s (top panel) and
the frustration ratio $\alpha$ (bottom panel) as a function of $U_{d}$ for $d_{xz}$ and 
$d_{yz}$ orbitals. The NN total exchange $J_{1}$ (full symbols) and NNN total exchange $J_{2}$
(empty symbols) do not
vary much for the considered range of $U_{d}$ (2.5 eV to 4.5 eV) values. The experimental
results (Ref. \onlinecite{nilsen08}) are indicated by dashed lines.  } 
\end{figure}

\begin{figure*}[t]
\begin{center}$
\begin{array}{ccc}
\includegraphics[width=2.0in]{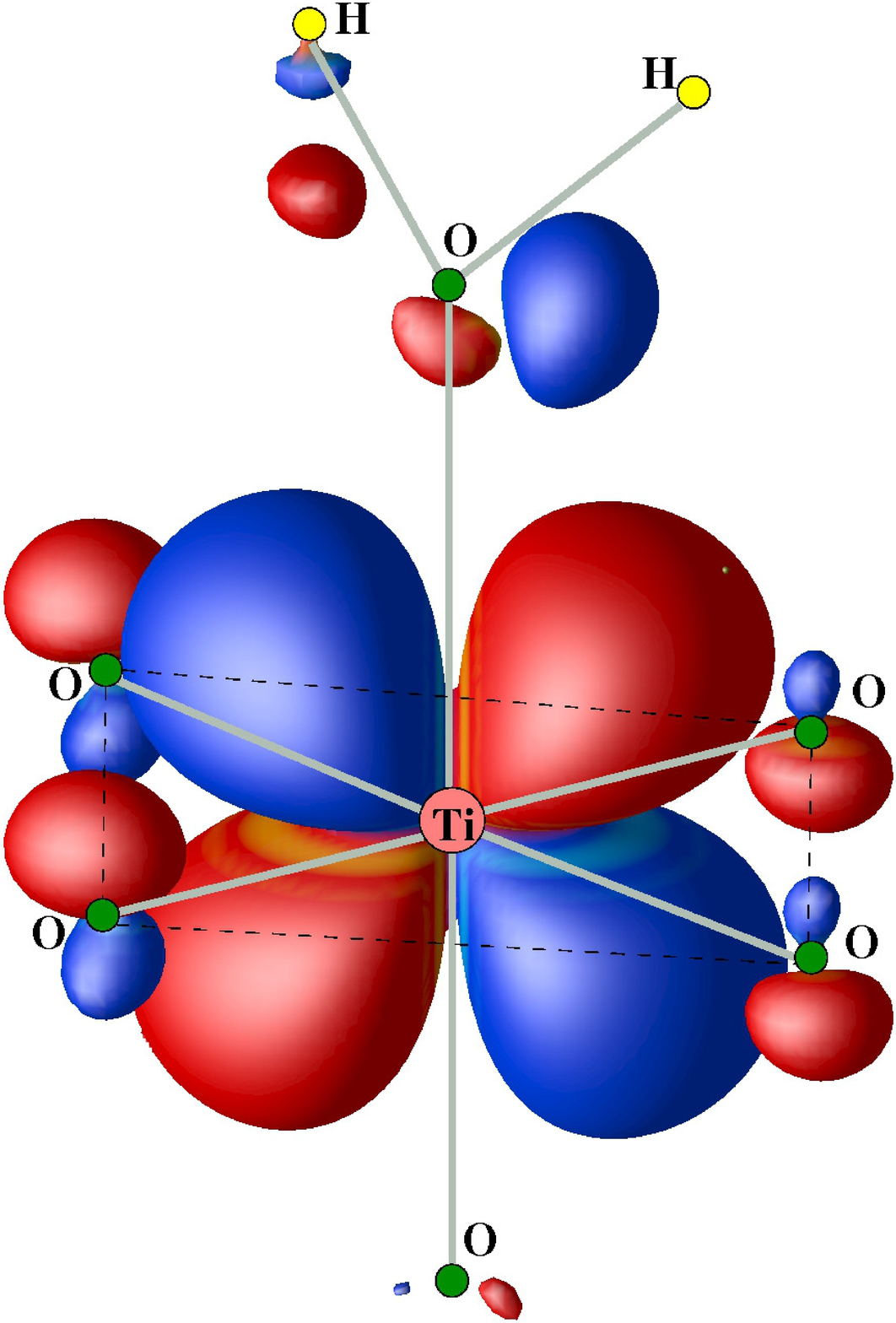} &
\includegraphics[width=2.0in]{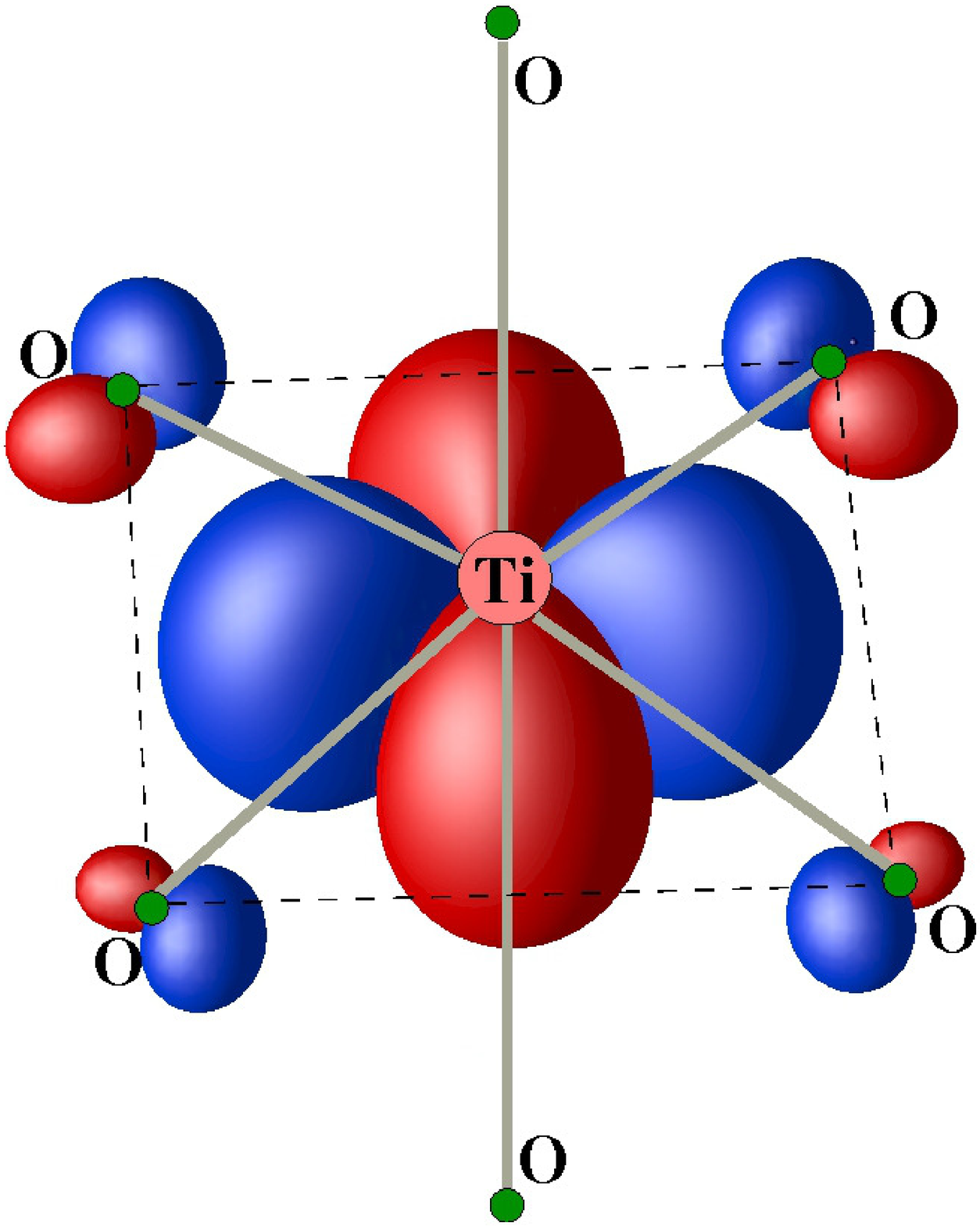} &
\includegraphics[width=2.0in]{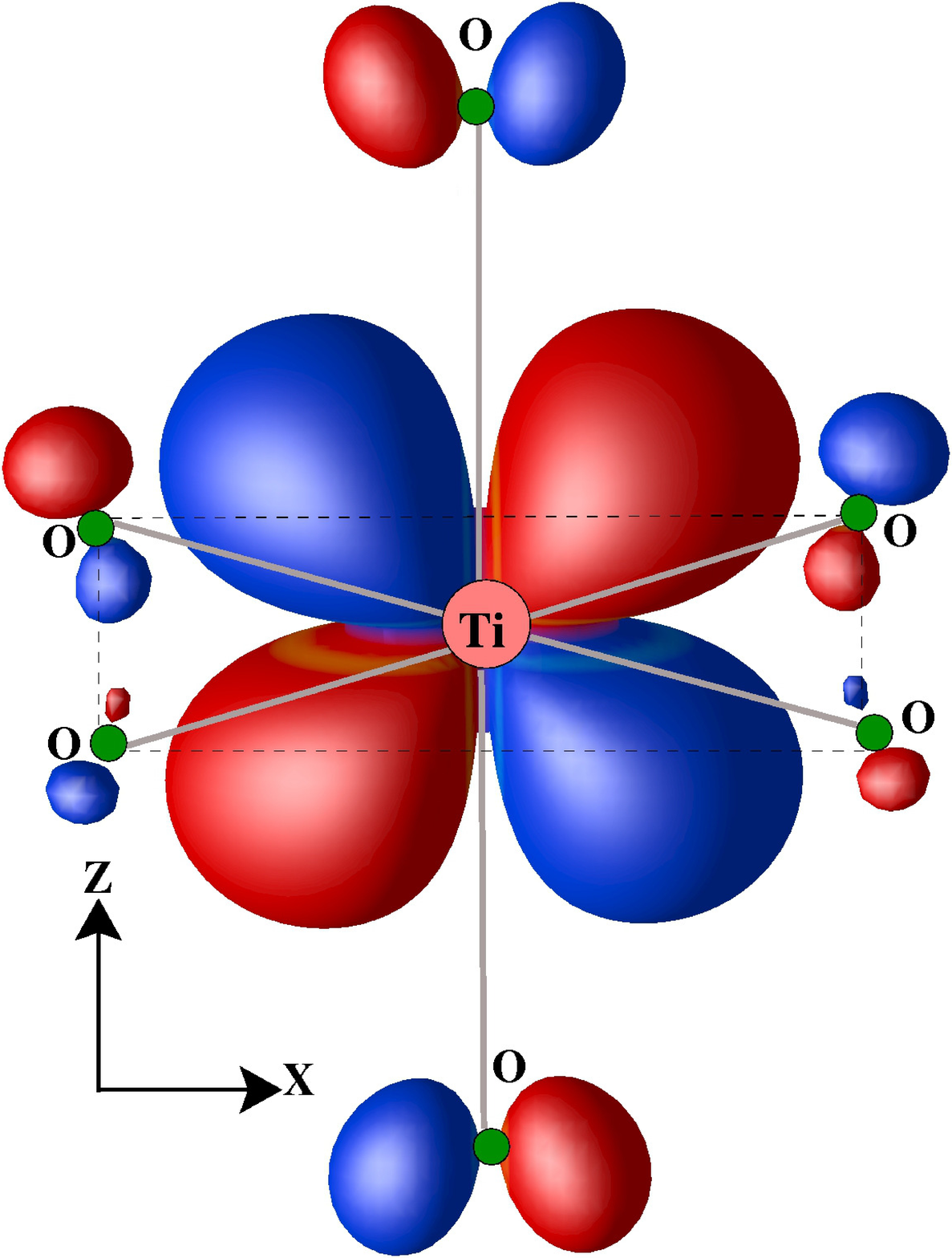} 
\end{array}$
\end{center}
\caption{\label{wan}(Color Online) Wannier functions of the three t$_{2g}$
orbitals of the Ti$^{3+}$ ion. {\bf Left panel:} d$_{xz}$, {\bf middle panel:} d$_{xy}$,
{\bf right panel:} d$_{yz}$. The blue and red lobes of the WF's refer to the 
positive and negative iso-surfaces, the circles refer to the atoms.
The TiO$_{6}$ octahedra is highligted by the light grey Ti-O bonds. The 
perspective view is also aided by highlighting the crystallograhpic $a-b$
plane using dashed lines. }
\end{figure*}

Since the constructed Heisenberg Hamiltonian takes into account only
the antiferromagnetic component of the couplings, inclusion of 
ferromagnetic (FM) component could
change the picture described above, especially considering the
absolute values for the AFM exchanges are quite small.
In the proceeding section we report on the LSDA+$U_{d}$ calculations
to obtain the total exchange constants $J$ = $J^{\mathrm{FM}} + J^{\mathrm{AFM}}$.

\subsection{Density functional theory + $U$}
Besides obtaining estimates
for all the AFM couplings in the system, the TBM also allows for
approximating the number of Ti-Ti neighbors that needs to be considered
when performing the more involved and time consuming LSDA+$U_{d}$ supercell 
calculations. 
Since the AFM exchanges obtained from the TBM beyond the NNN are 
less than 0.1 meV, and because FM interactions beyond second neighbors should also
be small, we
constructed two supercells to obtain the values of the
short-ranged exchanges $J_{\mathrm{1}}$ and 
$J_{\mathrm{2}}$.
Using different initial density matrices for the Ti 3$d$ orbitals, one can
correlate (fill the spin-up band with one electron and leave the spin-down
band empty) the bands belonging to different irreducible representations. 
For  KTi(SO$_{4})_{2}\cdot$H$_{2}$O we tried to spin-polarize 
each of the three $t_{2g}$ bands. We considered $U_{d}$ values ranging from
2.5 - 4.5 eV.\cite{J0} First let us consider the $d_{xz}$ and $d_{yz}$ orbitals, which
gave similar AFM exchange constants in our TBM. In all of our LSDA+$U_{d}$ 
calculations, the scenario in which the $d_{xz}$ band
was spin-polarized had the lowest energy. Spin-polarizing $d_{yz}$ required an 
additional energy of 350 meV per Ti$^{3+}$ ion. This energy scale is comparable
to the band width of these orbitals. Incidentally, all our attempts to 
spin-polarize $d_{xy}$ remained unsuccessful, though not unexpected 
when considering our TBM results ($i.e.$ the much too large energy scale of $J_{2}$ which is inconsistent with experiments). 
We were also unable to stabilize different orbitally-ordered
scenarios ($i.e.$ one Ti ion with a spin-polarized $d_{xz}$ orbital and the
NN Ti ion with a spin-polarized $d_{yz}$ orbital).
This alludes to the fact that the (local) magnetic ground state in
 KTi(SO$_{4})_{2}\cdot$H$_{2}$O  is very likely determined by the Ti 3$d_{xz}$ 
orbital.  The next question to answer is whether the exchange constants
obtained for the $d_{xz}$ orbital are consistent with the experimental
findings. 
We obtain effective
exchange constants by performing LSDA+$U_{d}$ calculations of differently
ordered spin configurations (FM, AFM and ferrimagnetic) and map the energies to a
Heisenberg model. Among the considered spin configurations,
the AFM spin configuration was always more favorable (lower total energy) than the FM. 
The exchange constants and the frustration ratio 
$\alpha$ are collected in Fig.~\ref{alpha}. 
For comparison, we have 
displayed the values for both $d_{xz}$ and $d_{yz}$ orbitals. 
For the range of $U_{d}$ values considered here, $J_{1}^{d_{xz}}$ is 
comparable to experimental findings while $J_{2}^{d_{xz}}$
is larger by almost an order of magnitude. Comparing the total $J$'s in 
Fig.~\ref{alpha} with the $J^{\mathrm{AFM}}$ obtained from TBM in Table.~\ref{tbm}, 
we can infer that there is a significant FM component to the first neighbor ($J_{1}^{\mathrm{FM}}$) 
while the FM componet to the NNN ($J_{2}^{\mathrm{FM}}$) is quite negligible.  
Though the $J$'s do not vary very
much for $U_{d}$ = 2.5 to 4.5 eV, an appropriate $U_{d}$ value needs to be chosen for
comparison with experiments.  
Spin- and orbital-unrestricted Hartree-Fock calculation of the on-site Coulomb interaction for various 
transition-metal oxides, recommend a $U_{d}$ value of 4 eV for Ti$^{3+}$ ions.\cite{fujimori} 
 Using that value
of $U_{d}$ as a benchmark, we obtain, J$_{1}^{d_{xz}}$ $\approx$ 12 K, J$_{2}^{d_{xz}}$ $\approx$
13.4 K and $\alpha_{xz}$ $\approx$ 0.94 (J$_{1}^{exp}$ = 9.46 K, J$_{2}^{exp}$ = 2.8 K,
and $\alpha$ = 0.29). 
The calculated value of $\alpha_{xz}$ = $J_{2}^{xz}/J_{1}^{xz}$ is larger than the experimental value by a factor of 4 for 
the $d_{xz}$ orbital. 
The calculated $\alpha_{yz}$ (= 0.60) for the energetically unfavorable $d_{yz}$ orbital is 
also somewhat larger than the experimental value. 
Albeit the S=1/2 frustrated chain magnetism in  KTi(SO$_{4})_{2}\cdot$H$_{2}$O
is  established in both LDA and LSDA+$U_{d}$ calculations, our results for $\alpha$ 
are not consistent with the 
experiments. Nonetheless, both our calculation and experiment suggest an $\alpha$ in the highly 
interesting region ( 0.2411 $<$ $\alpha$ $<$ 1.8) of the spin-1/2 frustrated 
chain phase diagram. 

In the following section, we attempt to understand the discrepancy between
our calculations and the experiment.

 \begin{figure*}[t]
\begin{center}$
\begin{array}{cc}
\includegraphics[width=3.0in]{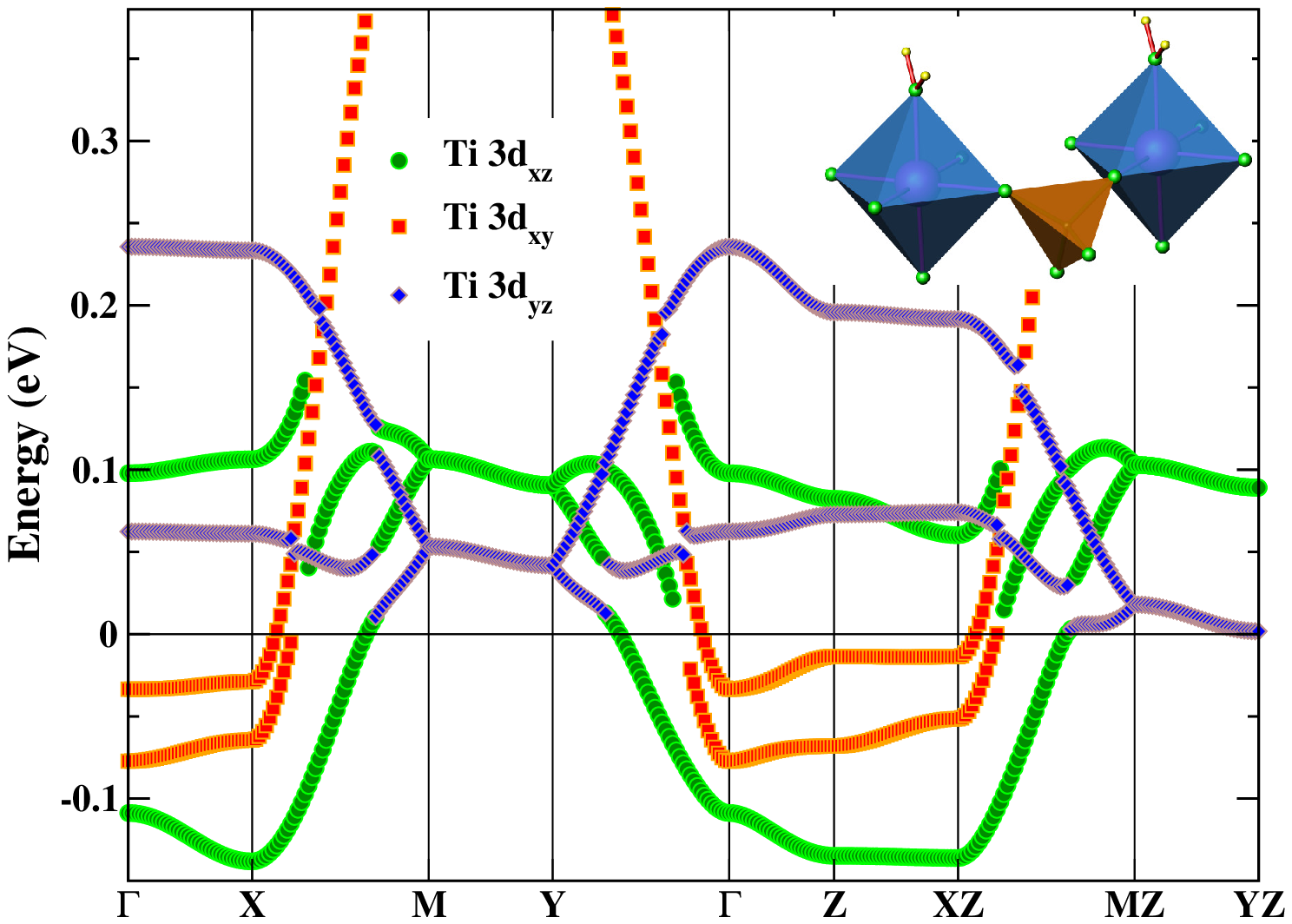} &
\includegraphics[width=3.0in]{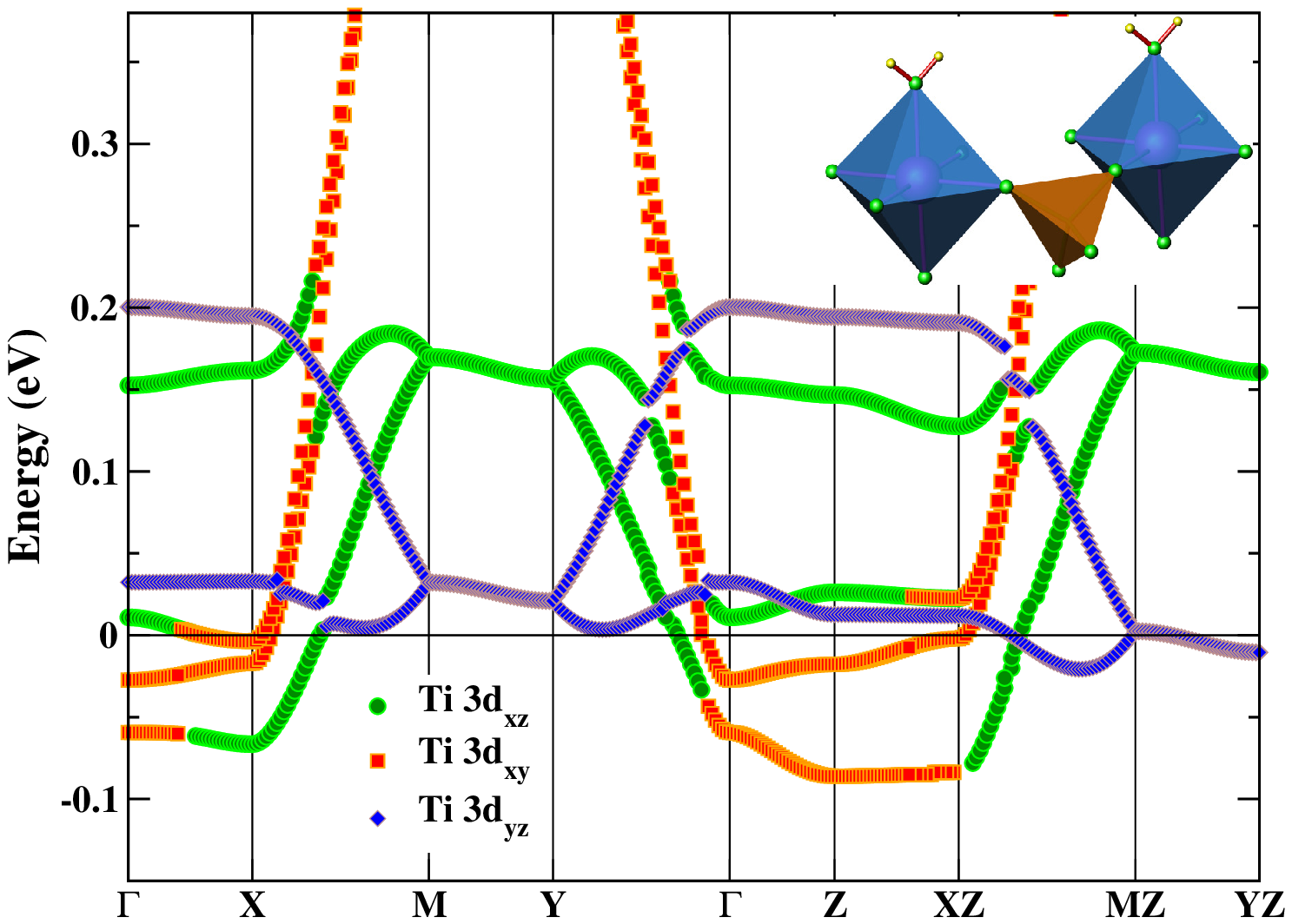} 
\end{array}$
\end{center}
\caption{\label{rot}(Color Online) Comparison of the non-magnetic band structure
for
KTi(SO$_{4})_{2}\cdot$H$_{2}$O using the reported $xz$-plane oriented (left) and the 
fictitious $yz$-plane oriented (right) H$_{2}$O molecule. The insets show a
 small segment of the double chains with the two respective orientations of the 
 H$_{2}$O molecule.   }
\end{figure*}

\section{Effect of crystal water}

\subsection{Wannier Functions}
The octahedron surrounding Ti$^{3+}$ consists of 5 O$^{2-}$ ions shared with SO$_{4}^{2-}$ groups, 
and one water molecule (see Sec. II for a more detailed discussion). From a naive consideration 
of average equatorial and apical Ti-O distances, the magnetically active orbital would be assigned as $d_{xy}$.
In contrast, both LDA and LSDA+$U_{d}$ results point to 
$d_{xz}$ as the magnetically active orbital. In order to shed some more 
light on this discrepancy, we have plotted the Wannier functions for
the Ti $t_{2g}$ complex (Fig.~\ref{wan}). 
Wannier functions are essentially a real-space picture of 
localized orbitals and can be used to enhance the understanding
of bonding properties via an analysis of factors such as their shape and symmetry.
Before analyzing the WF's, one should keep in mind that the two nearest neighbor (NN) Ti
atoms do not belong to the same chain but to the pair-chain displaced along
the $c$-axis (see lower
panel of Fig.~\ref{str}). By fitting (using exact diagonalization) the low
temperature magnetic susceptibility, the recent experimental report\cite{nilsen08} suggests that
the AF-NN interaction $J_{1}$ is larger than the
AF-NNN interaction $J_{2}$. Therefore, one expects to observe large tails 
at oxygen sites from the WF's bending towards the NN Ti atoms, as this
facilitates	 the Ti-O-O-Ti superexchange.  
The Ti $d_{xy}$ WF is composed of contributions from the $d_{xy}$ 
orbital as well as tails on the oxygen sites in the $xy$-plane, although these do not point towards the NN Ti atom.
The $d_{yz}$  WF, on the otehr hand, has tails on all 6 oxygen sites of the TiO$_{6}$ octahedra,
and all point towards the NN Ti atom.
Interestingly, the $d_{xz}$ WF not only has tails on the oxygen sites bending towards the NN Ti, but also has tails on one of the hydrogen site belonging to the crystal water
molecule. Such an effect is arising from the hybridization effect of the O and H orbitals.
The effects of hybridization involving the H-atom in the $d_{yz}$ WF is comparatively less than the $d_{xz}$, since no extended tails are observed on the H site.   
We therefore infer that the orientation of the water molecule plays
an important role in deciding the magnetically active orbital for 
obtaining the ground state in  KTi(SO$_{4})_{2}\cdot$H$_{2}$O. More
detailed analysis follows in the  next section.

\subsection{Rotating the water molecule}

It is normally assumed that the presence of crystal water in a 
compound, leads to only modest
changes in the crystal field of the magnetic ion. However, it might
not always be the case. Recent work \cite{cucl2} on CuCl$_{2}$ and CuCl$_{2}\cdot$2H$_{2}$O 
shows that the former is a quasi 1D-chain compound while the latter 
is a classical 3D AFM, although the magnetic Cu-Cl units in 
both compounds are structurally similar. Motivated by such reports and the
results from our WF analysis, we explore the effects of the
orientation of the H$_{2}$O molecule in KTi(SO$_{4})_{2}\cdot$H$_{2}$O.
Crystallographic data that we have used until now orient the 
H$_{2}$O molecule in the $xz$ plane. From the analysis of the WF's we infer that
the magnetically active orbital ($d_{xz}$) is selected by the hydrogen bonding 
of the water molecule in the $xz$-plane
with the Ti $d_{xz}$ orbital. 
We now attempt to rotate the H$_{2}$O molecule to
the $yz$ plane and observe the effects on the electronic structure.
The symmetry of the crystal was reduced to accommodate the new
orientation, though the experimental O-H distance (0.8747 \AA) as well as the H-O-H bond 
angle (93.33$^{\degree}$) were left untouched. The resulting 
non-magnetic LDA band structure is displayed in Fig.~\ref{rot}. 
The general shape of the bands as well as the various $k$ - point
degeneracies are similar for both the $xz$ and $yz$ oriented
H$_{2}$O molecule. The important changes for the fictitious $yz$ oriented 
system are:
(a) the $d_{xz}$ band is no longer having the lowest band energy and
is getting pushed upwards and therefore less occupied, 
(b) the $d_{yz}$ band is pulled 
downwards and is getting more occupied. 
Ergo, the orientation of the H$_{2}$O molecule has a profound
effect on the electronic structure and plays an important role in
selecting the magnetically active orbital.
In contrast with the LDA result, the addition of correlations results in $d_{xz}$ as the lowest 
energy band, as was the case for H$_{2}$O || $xz$. That said, the energy difference between 
singly occupied $d_{xz}$ and $d_{yz}$ is only 35 meV, 10 times smaller than the original calculation (cf. Sec.~IV-C) in which the H$_{2}$O molecule was
oriented along the $xz$-plane. 
Calculating the $J$'s for the
fictitious system using the effective TBM resulted in $\alpha_{yz}$ = 0.36 
and a much larger $\alpha_{xz}$ = 1.11.
These results reaffirm the the importance of the 
water molecule in deciding not only the magnetically 
active orbital, but also the accordingly derived exchanges and frustration ratio in
KTi(SO$_{4}$)$\cdot$H$_{2}$O.

\subsection{Effect of the O-H bond length}
Having observed the sensitivity of the electronic structure
to the orientation of the H$_{2}$O molecule, our next item of interest was
to check the importance of the O-H bond length on the electronic structure and magnetic properties.
Recent reports on spin-1/2 kagome lattice systems \cite{kagome} show a dramatic
impact of the O-H bond length on the exchange. 
It is well known that
obtaining the correct O-H bond length via X-ray diffraction in a system containing heavy atoms is difficult.
OH$^{-}$ impurity in oxides
have been shown to have an equilibrium bond length of $\approx$ 1~\AA.\cite{o-hbond}
For KTi(SO$_{4})_{2}\cdot$H$_{2}$O, the reported O-H bond length is
10\% smaller and 
only 0.874~\AA. One possible reason for the larger frustration ratio $\alpha$
in LSDA+$U_{d}$ calculations as compared to experiments might arise from
the possibly underestimated O-H bond length.\cite{hydrogen} We have therefore allowed the O-H bond length
to relax.
Keeping the TiO$_{6}$ octahedra rigid, we relaxed
the H position with respect to the total energy and obtained
an optimized O-H bond length of about 1~\AA, in accordance with the empirical expectations.\cite{o-hbond,kagome}  
Recalculating the exchange constants using the 
optimized O-H distance, 
we obtain for $U_d$ = 4.0~eV, $J_{1}^{xz}$ = 10~K, $J_{2}^{xz}$ = 14.2~K.
The frustration ratio $\alpha_{xz}$ = 1.4 is even larger than the previously
calculated value using the experimental O-H bond length.  
Thus far, all of our calculations have resulted in exchanges that have the correct
sign and comparable magnitudes to the previous experimental report, 
but nonetheless always yielded frustration ratios far greater than previously reported.\cite{nilsen08}.
In the following section, we make a final attempt to clarify the 
discrepancy between the experiment and our calculations.

\section{TMRG calculations}

\begin{figure}[tb]
\begin{center}
\includegraphics[%
  clip,
  width=8.5cm,
  angle=-0]{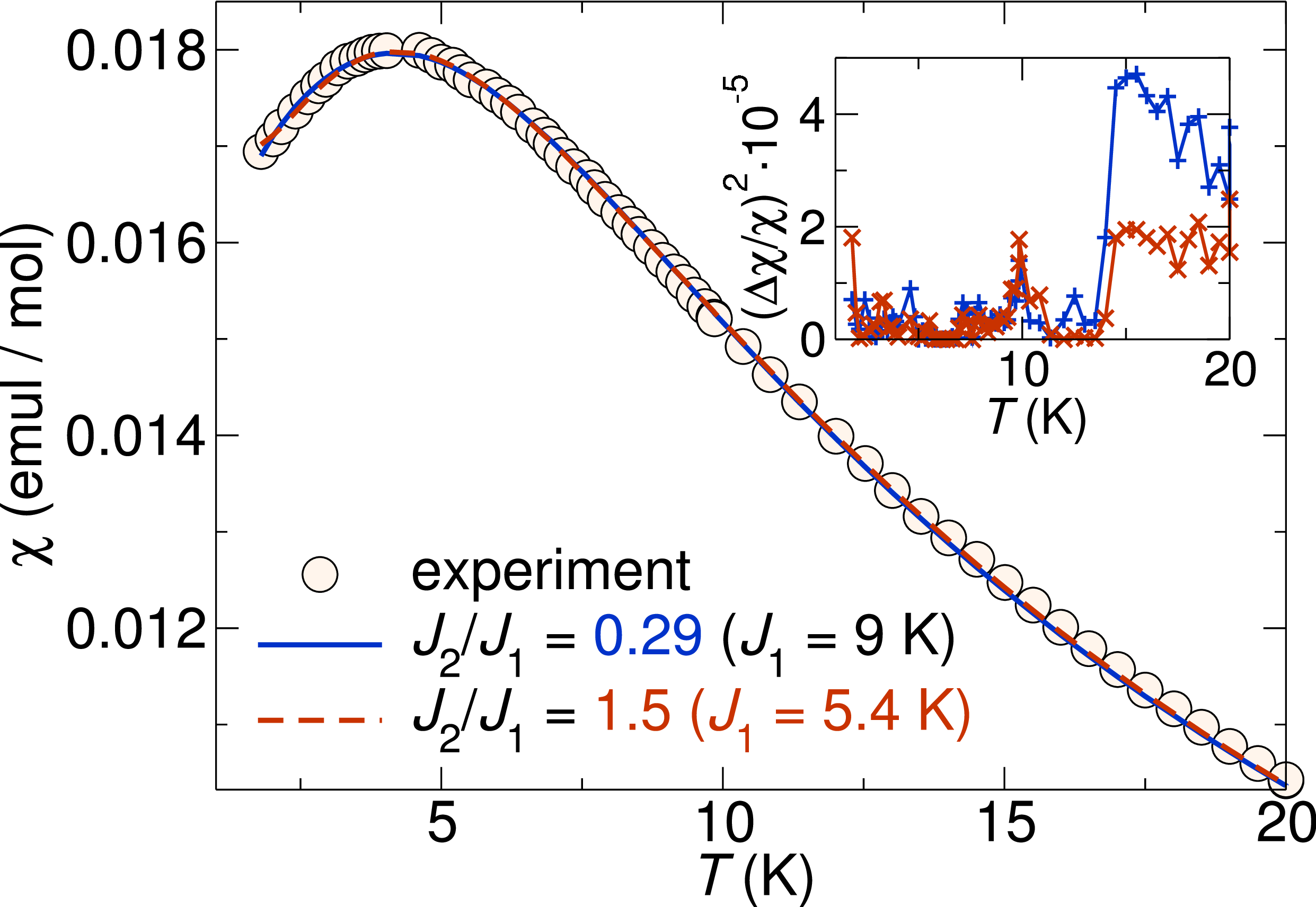}
\end{center}
\caption{\label{F-chi}(Color Online) Fits to the
magnetic susceptibility. The experimental data are adopted from
Ref.~\onlinecite{nilsen08}. Magnetic susceptibility of frustrated Heisenberg chains
with $\alpha\!=\!0.29$ and $\alpha\!=\!1.5$ ($\alpha\!\equiv\!J_2/J_1$) was
simulated using TMRG. The simulated curves were fitted to the experiment by
varying the fitting parameters $J_1$, $g$ and the temperature-independent
contribution $\chi_0$. Inset: difference curves for both solutions.
$\Delta\chi$ is the difference between the simulated and the experimental
value.} 
\end{figure}

\begin{figure}[tb]
\begin{center}
\includegraphics[%
  clip,
  width=9cm,
  angle=-0]{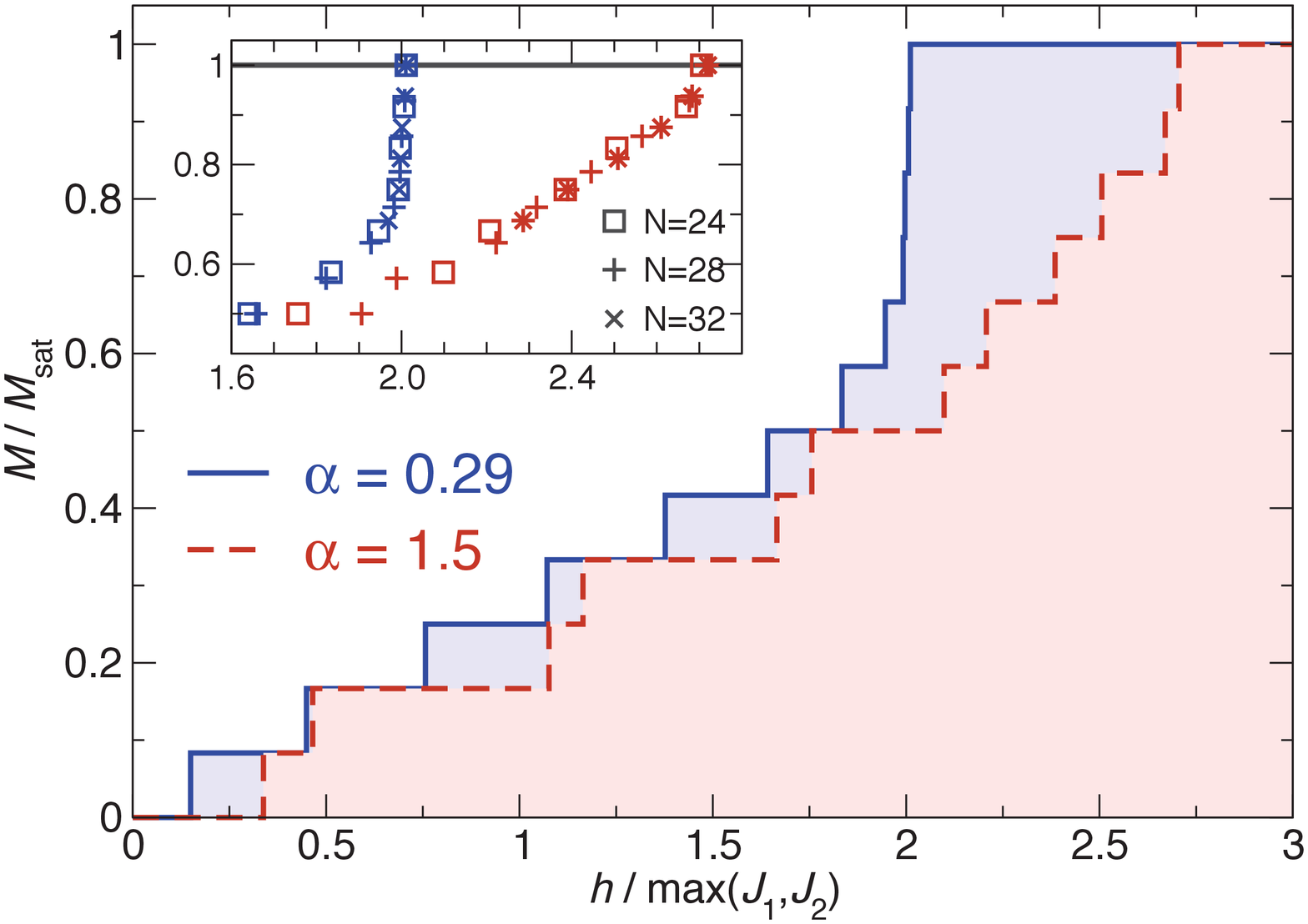}
\end{center}
\caption{\label{F-MH}(Color Online) Ground state magnetization of frustrated
Heisenberg chains with $\alpha\!=\!0.29$ and $\alpha\!=\!J_2/J_1\!=\!1.5$
($\alpha\!\equiv\!J_2/J_1$), simulated using exact diagonalization on finite
lattices (rings) of $N$\,=\,24 spins. Note the characteristic upward bending of
the $\alpha\!=\!0.29$ curve. Inset: finite-size dependence of the ground state
magnetization.}
\end{figure}

As demonstrated in Ref.~\onlinecite{nilsen08}, a frustrated Heisenberg
chain model with $\alpha\!=\!0.29$ can reproduce the experimental
magnetic susceptibility curve of KTi(SO$_{4})_{2}\cdot$H$_{2}$O. The
small $\alpha\!=\!0.29$ implies that $J_1$ is large, while $J_2$ is
small. This is at odds with our LSDA+$U_d$ calculations, where the
antiferromagnetic exchange between NNN Ti atoms appears to be more
efficient than the NN exchange.  Thus, in the microscopic model,
$\alpha$ should be larger than unity. The question is then, whether the
large $\alpha$ regime conforms to the experimental behavior.

The small energy scale of the leading couplings leads to sizable error
bars for the $J_1$ and $J_2$ values estimated from LSDA+$U_d$
calculations.  To refine the model parameters, we use the analytical
expressions for the high-temperature part of the magnetic susceptibility
of a frustrated Heisenberg chain, the high-temperature series expansion
(HTSE).\cite{buehler01}  Typical for a local optimization procedure, the
results are dependent on the initial values.  If we start from the
$J_1>J_2$ limit, HTSE yields $J_1\!\simeq\!9.6$ K and $J_2\!\simeq\!2.9$ K,
very close to the $\alpha\!=\!0.29$ reported in
Ref.~\onlinecite{nilsen08}. In contrast, if we proceed from the
$J_2>J_1$ regime, we obtain $J_1\!\simeq\!5.4$ K and $J_2\!\simeq\!8.1$ K ($\alpha$ = 1.5),
in accord with our LSDA+$U_d$ calculations. Thus, HTSE yields two 
ambiguous solutions.

HTSE typically diverges at temperatures comparable with the magnetic
energy scale ($T\geq{}J$).  To verify, whether both solutions agree with
the experimental $\chi(T)$ at lower temperatures, we simulate the
temperature dependence of reduced magnetic susceptibility $\chi*$ using
TMRG, and fit the resulting $\chi^*(T/\max{\{J_1,J_2\}})$ dependencies
to the experimental curve. In this way, we again find that besides the
previously reported $\alpha\,=\,0.29$ solution, the $\alpha$\,=\,1.5
curve with $J_1$\,=\,5.4\,K, $J_2$\,=\,8.1\,K, $g$\,=\,1.74, and
$\chi_0\,=\,5.9{\cdot}10^{-5}$\,emu/mol also yields an excellent fit to
the experimental magnetic susceptibility (Fig.~\ref{F-chi}).  The
difference curves evidence that both $\alpha$\,=\,0.29 and
$\alpha$\,=\,1.5 provide equally good description of the experimental
$\chi(T)$ data (Fig.~\ref{F-chi}, inset).

The coexistence of the two solutions actually manifests the inner symmetry of the
frustrated chain model. A trivial example: the uniform chain limit can be
described with $\alpha=+0$ ($J_1\neq 0$, $J_2\!=\!0$) as well as
$\alpha\!=\!\infty$ ($J_1\!=\!0$, $J_2 \neq 0$).  The $\alpha$\,=\,0.29
and $\alpha$\,=\,1.5 solutions are also related, although in a less
trivial way. To pinpoint this relation, we revisit the phase diagram of
the frustrated chain model. The $\alpha\,=\,0$ limit corresponds to the
exactly solvable gapless Heisenberg chain model. This GS is robust
against small frustrating $J_2$, up to the quantum critical point
$\alpha_{c}\,\simeq\,0.2411$, where a spin gap opens.\cite{okamoto}  For
larger $\alpha$ values, the spin gap rapidly increases and reaches its
maximum value $\Delta\,\simeq\,0.43$\,$J_1$ at $\alpha\,\simeq\,0.6$.
Further enhancement of $\alpha$ reduces the spin gap. In the large
$\alpha$ limit ($J_2\,\gg\,J_1$), the spin gap exhibits an exponential
decay.\cite{White}

Thus, for a certain value of the spin gap $\Delta$, there are two
possible $\alpha$ values: (i) with a dominant $J_1$, i.e.\ from the
$\alpha$\,=\,0.2411--0.6 range, and (ii) with a sizable $J_2$
($\alpha$\,=\,0.6--$\infty$). Since $\Delta$ plays a decisive role for the
shape of $\chi(T)$, both solutions yield similar macroscopic magnetic
behavior. This explains the seemingly unusual fact that the experimental
data for KTi(SO$_{4})_{2}\cdot$H$_{2}$O can be fitted by both
$\alpha\!=\!0.29$ and $\alpha\!=\!1.5$.

Despite the similar spin gaps, the solutions $\alpha$\,=\,0.29 and
$\alpha$\,=\,1.5 are physically different (unlike, e.g.\, $\alpha$\,=\,0
and $\alpha$\,=\,$\infty$, that describe the same physics): 
spiral correlations are present in the latter case,
only.\cite{White} Moreover, the two solutions feature substantially
different correlation lengths.\cite{White}
 Thus, the two solutions can be
distinguished by  measuring a characteristic experimental feature
(``smoking gun'').

For spin systems, a measurement of magnetization isotherms is 
technically simple, but very informative, especially for
systems with weak magnetic couplings.  Since the magnetic field linearly
couples to the $S^z$ component of the spin, a magnetization curve
reflects the energy of the lowest lying state in each $S^z$ sector.  This
often suffices to distinguish between ambiguous solutions.  For
instance, HTSE for the $J_1-J_2$ square lattice system
BaCdVO(PO$_4$)$_2$ yielded, besides the frustrated solution with an AFM
$J_2$, also a nonfrustrated solution with AFM $J_2$. However, the
frustrated scenario was clearly underpinned by $M(H)$
measurements.\cite{nath08} In a recent work, $M(H)$ measurements for
$A_2$CuP$_2$O$_7$ ($A$\,=\,Li,Na) resolved previous controversies
concerning the magnetic dimensionality of these
compounds.\cite{lebernegg11}

We argue that for the frustrated Heisenberg chain model, the
characteristic behavior of magnetization on the verge of saturation can
be used to distinguish between different scenarios.  In particular, the
$\alpha\!=\!0.29$ magnetization curve exhibits a well pronounced upward
bending, while only a feeble bending is visible in the $\alpha\,=\,1.5$
GS magnetization (Fig.~\ref{F-MH}). Another relevant quantity is the
saturation field:
\begin{equation}
\label{E-Hsat} H_{\text{sat}} =
(g\mu_{\text{B}})^{-1}\left[E(S^z_{\text{max}}) - E(S^z_{\text{max}}-1)\right],
\end{equation}
where $S^z_{\text{max}}$ corresponds to the fully polarized state.  The
energies are estimated using exact diagonalization for finite chains of
$N$\,=\,32 spins.  Adopting $J_1$, $J_2$, and $g$ values from the HTSE
fits, we obtain $H_{\text{sat}}\!\simeq\!16.4$\,T and
$H_{\text{sat}}\!\simeq\!18.7$\,T for $\alpha\!=\!0.29$ and
$\alpha\!=\!1.5$, respectively.  Both values of saturation field lie in
the experimentally accessible field range. A somewhat problematic point
could be the low energy scale of KTi(SO$_{4})_{2}\cdot$H$_{2}$O, which
renders the typical measurement temperature of $\sim$1.5\,K as
relatively high, hence the states with different $S^z$ could be
substantially mixed.  Still, the $\alpha\!=\!0.29$ magnetization
isotherms will retain fingerprints of the characteristic bending.
Therefore, we believe that a high-field (up to $\sim$20\,T) measurement
of a magnetization isotherm will be an instructive and decisive
experiment to distinguish between the $\alpha\!=\!0.29$ and
$\alpha\!=\!1.5$ scenarios.

\section{Summary}
In conclusion, we have studied the electronic structure of KTi(SO$_{4})_{2}\cdot$H$_{2}$O
in detail using DFT based calculations. The results of 
both the TBM and LSDA+$U_{d}$ calculations confirm beyond doubt the low
dimensional nature of the material with NN and NNN exchanges $J_{1}$ and
$J_{2}$ confined to the double-chains running along the $b$-axis. 
We also confirm the AFM nature of the exchanges, consistent with the
experimental report, with the Ti-3$d_{xz}$ orbital being the magnetically 
active one, holding the single unpaired electron of the Ti$^{3+}$ ion.
The magnitude of the calculated $J$'s are of the 
right order compared to the experiment, though we observe a strong
dependence to the $t_{2g}$ orbital choice. 
Notwithstanding the small energy scale ($\approx$ 10 K) of the system, we are
able to obtain accurate results from our DFT calculations. 
Additionally, we observe a 
surprisingly large spin-lattice coupling with respect to the crystal water related 
degrees of freedom, which in turn manifests itself by playing an 
important role in determining the ground state of the system. 
This feature is clearly elucidated by calculating the Wannier functions,
which show the effects of hydrogen bonding to the corresponding 
$t_{2g}$ orbital which is oriented in the same plane as the crystal 
water molecule. 
Using the experimental position for hydrogen, we obtain a frustration 
ratio $\alpha \approx$ 1.0 and a value of $\alpha \approx$ 1.4 upon 
relaxing the hydrogen position in the crystal lattice. 
Both these values are larger than the experimental value $\alpha_{\mathrm{exp}}$ = 0.29. 
In order to understand the origin of this discrepancy between the experiment 
and our calculations, we simulated the temperature dependence of the susceptibility 
using both the small and large values of $\alpha$. 
Due to an intrinsic symmetry of the $J_{1}-J_{2}$ frustrated chain model, we show that
both values of $\alpha$ provide equally good fits to the experimental curve. 
Consequently, we calculated magnetization curves as a means to unambiguously 
distinguish the two solutions and show two features which can be used to identify the
appropriate $\alpha$ that defines the magnetic ground state of KTi(SO$_{4})_{2}\cdot$H$_{2}$O. 
Hence, we suggest performing high-field magnetization measurements on this 
system and as well as a.c. susceptibility experiments (to obtain the size of the spin-gap)
at very low temperatures.


\begin{thebibliography}{99}

\bibitem{DrechslerEPL} W.~E.~A.~Lorenz, R.~O.~Kuzian, S.-L.~Drechsler, W.-D.~Stein, N.~Wizent, G.~Behr, 
J.~M\'{a}lek, U.~Nitzsche, H.~Rosner, A.~Hiess, W.~Schmidt, R.~Klingeler, M.~Loewenhaupt, and B.~B\"{u}chner,
Europhys. Lett. {\bf 88}, 37002 (2009).

\bibitem{Nitzsche} U.~Nitzsche, S.-L.~Drechsler and H.~Rosner, to be published.


\bibitem{Gippi04} A.~A.~Gippius, E.~N.~Morozova, A.~S.~Moskvin, A.~V.~Zalessky, A.~A.~Bush,
M.~Baenitz, H.~Rosner, and S.-L.~Drechsler, Phys. Rev. B {\bf 70}, 020406 (2004).

\bibitem{Drechs05} S.-L.~Drechsler, J.~M\'{a}lek, J.~Richter, A.~S.~Moskvin, A.~A.~Gippius, H.~Rosner,
Phys. Rev. Lett. {\bf 94}, 039705 (2005).

\bibitem{Enderle} M.~Enderle, C.~Mukherjee, B.~Fak, R.~Kremer, J.~Broto, H.~Rosner, S.-L.~Drechsler,
J.~Richter, J.~M\'{a}lek, A.~Prokofiev, et~al., Europhys. Lett.
{\bf 70}, 237 (2005).

\bibitem{Masuda} T.~Masuda, A.~Zheludev, A.~Bush, M.~Markina, and A.~Vasiliev, Phys. Rev. Lett. {\bf 92}, 177201 (2004).

\bibitem{Capogna} L.~Capogna, M.~Mayr, P.~Horsch, M.~Raichle, R.~Kremer, M.~Sonin, and B.~Keimer, Phys. Rev. B {\bf 71}, 140402 (2005).

\bibitem{Drechs06} S.-L.~Drechsler, J.~Richter, A.~Gippius, A.~Vasiliev, A.~Bush, A.~Moskvin, Y.~Prots, W.~Schnelle, 
and H.~Rosner, Europhys. Lett. {\bf 73}, 83 (2006).


\bibitem{LiCu2O2} V.~V.~Mazurenko, S.~L.~Skornyakov, A.~V.~Kozhevnikov, F.~Mila, and V.~I.~Anisimov, Phys. Rev. B {\bf 75}, 224404 (2007).



\bibitem{CuGeO3} J.~Riera, and A.~Dobry, Phys. Rev. B {\bf 51}, 16098 (1995).

\bibitem{hase}M.~Hase, I.~Terasaki, and K.~Uchinokura, Phys. Rev. Lett. {\bf 70}, 3651 (1993).

\bibitem{LaTiO3}G.~ Khaliullin, and S.~ Maekawa, Phys. Rev. Lett. {\bf 85}, 3950 (2000).

\bibitem{YTiO3}T.~Kiyama, H.~Saitoh, M.~Itoh, K.~Kodama, H.~Ichikawa, J. ~Akimitsu, J. Phys. Soc. Jpn. {\bf 74}, 1123 (2005). 


\bibitem{seidel} A.~Seidel, C.~A.~Marianetti, F.~C.~Chou, G.~Ceder, and P.~A.~Lee, Phys. Rev. B {\bf 67}, 020405(R) (2003).

\bibitem{nilsen08} G.~J.~Nilsen, H.~M.~R{\o}nnow, A.~M.~L\"{a}uchli, F.~P.~A.~Fabbiani, J.~Sanchez-Benitez, K.~V.~Kamenev, and 
A.~Harrison, Chem. Mater. {\bf 20}, 8 (2008).



\bibitem{alpha1}R.~Jullien, and F.~D.~M. Haldane, Bull. Am. Phys. Soc. {\bf 28}, 344 (1983).

\bibitem{alpha2}K.~Okamoto, and K.~Nomura, Phys. Lett. A {\bf 169}, 433 (1992).

\bibitem{alpha3}S.~Eggert, Phys. Rev. B {\bf 54}, R9612 (1996).

\bibitem{alpha4}C.~K.~Majumdar, and D.~K.~Ghosh, J. Math. Phys. {\bf 10}, 1388 (1969).

\bibitem{alpha5}C.~K. Majumdar, and D.~K. Ghosh, J. Math. Phys. {\bf 10}, 1399 (1969).

\bibitem{alpha6}F.~D.~M.~Haldane, Phys. Rev. B {\bf 25}, 4925 (1982).

\bibitem{alpha7}S.~R.~White and I.~Affleck, Phys. Rev. B {\bf 54}, 9862 (1996)

\bibitem{krausite} E.~J.~Graeber, B.~Morosin, and A.~Rosenzweig, American Mineralogist {\bf 50}, 1929 (1965).

\bibitem{fn}Note that, the orbital angular momentum remains 
unquenched in the 2T ground state term, transforming as $-\mathbf{L}$ for
$L=1$. Spin-orbit coupling results in a pair of non-magnetic Kramers doublets in the effective $J$ = 3/2 ground 
state and a magnetic Kramers doublet excited state. 
However, the spin-orbit coupling fir Ti is rather small compared to the ligand field 
split due to the small distortion of the TiO$_{6}$ octahedra. Thus, we do not 
include the spin-orbit coupling in our calculations.

\bibitem{saha} T.~Saha-Dasgupta, R.~Valenti, H.~Rosner, and C.~Gros, Europhys. Lett. {\bf 67}, 63 (2004).

\bibitem{fplo1} K.~Koepernik, and H.~Eschrig, Phys. Rev. B {\bf 59}, 1743 (1999).

\bibitem{fplo2} I.~Opahle, K.~Koepernik, and H.~Eschrig,  Phys. Rev. B {\bf 60}, 14035 (1999).

\bibitem{PW92} J.~P.~Perdew, and Y.~Wang, Phys. Rev. B {\bf 45}, 13244 (1992).

\bibitem{wanfun} H.~Eschrig, and K.~Koepernik, Phys. Rev. B {\bf 80}, 104503 (2009).


\bibitem{ALPS}A.~F.~Albuquerque, F.~Alet, P.~Corboz, P.~Dayal, A.~Feiguin, S.~Fuchs, L.~Gamper, E.~Gull, S.~G\"{u}rtler, A.~Honecker, R.~Igarashi, M.~K\"{o}rner, A.~Kozhevnikov, A.~L\"{a}uchli, S.~R.~Manmana, M.~Matsumoto, I.~P.~McCulloch, F.~Michel, R.~M.~Noack, G.~Pawlowski, L.~Pollet, T.~Pruschke, U.~Schollw\"{o}ck, S.~Todo, S.~Trebst, M.~Troyer, P.~Werner, and S.~Wessel, J. Magn. Magn. Mater. {\bf 310}, 1187 (2007).

\bibitem{tmrg}X.~Wang, and T.~Xiang, Phys. Rev. B {\bf 56}, 5061 (1997).

\bibitem{CuSbO}D.~Kasinathan, K.~Koepernik, and H.~Rosner, Phys. Rev. Lett. {\bf 100}, 237202 (2008).

\bibitem{J0}The effective onsite exchange $J_{\mathrm{d}}$ has been set to 1 eV.

\bibitem{fujimori}T.~Mizokawa, and A.~Fujimore, Phys. Rev. B {\bf 54}, 5368 (1996). One should note that the value of 
$U_{d}$ is depending on the basis set, the choice of the double counting scheme, etc. Thus, $U_{d}$ = 4 eV should 
be considered as an approximation, only.


\bibitem{cucl2} M.~Schmitt, O.~Janson, M.~Schmidt, S.~Hoffmann, W.~Schnelle, S.-L.~Drechsler, and H.~Rosner, Phys. Rev. B {\bf 79}, 245119 (2009).


\bibitem{kagome} O.~Janson, J.~Richter, and H.~Rosner, Phys. Rev. Lett. {\bf 101}, 106403 (2008).

\bibitem{o-hbond}V.~Szalay, L.~Kov\'{a}cs, M.~W\"{o}hlecke, and E.~Libowitzky, Chem. Phys. Lett. {\bf 354}, 56 (2002).

\bibitem{hydrogen}The electron density of hydrogen, the lightest atom with only one electron is generally localized away 
from the nucleus and hence it is
difficult to detect the exact position of the H nucleus from X-ray diffraction measurements. The relatively high electron density between
the oxygen and hydrogen atoms,  makes the O-H bonds to appear too short. 

\bibitem{buehler01} A.~B{\"{u}}hler, U.~L{\"{o}}w, and G.~Uhrig, Phys.~Rev.~B {\bf 64}, 024428 (2001).

\bibitem{okamoto}K.~Okamoto, and N.~Nomura, Phys. Lett. A {\bf 169}, 433 (1992).

\bibitem{White} S.~R.~White, and I.~Affleck, Phys. Rev. B. {\bf 54}, 9862 (1992).

\bibitem{nath08} R.~Nath, A.~A.~Tsirlin, C.~Geibel, and H.~Rosner, 
Phys.~Rev.~B {\bf 78}, 064422 (2008).


\bibitem{lebernegg11} S.~Lebernegg, A.~A.~Tsirlin, O.~Janson, R.~Nath,
J.~Sichelschmidt, {\mbox{Yu}}.~Skourski, G.~Amthauer, and H.~Rosner,
Phys.~Rev.~B {\bf 84}, 174436 (2011).





\end{thebibliography}
\end{document}